\documentclass[final,1p,times]{elsarticle}

\usepackage{subfigure}
\usepackage{amsmath}
\usepackage{tabularx}
\usepackage{float}
\usepackage{listings}
\usepackage{color}
\usepackage{textcomp}
\definecolor{listinggray}{gray}{0.9}
\definecolor{lbcolor}{rgb}{1,1,1}
\usepackage{graphicx}

\usepackage{amssymb}

\journal{Biomedical Signal Processing and Control}

\begin{document}

\begin{frontmatter}



\title{A new algorithm for wavelet-based heart rate variability analysis}

\author[citius]{Constantino~A.~Garc\'{i}a\corref{cor1}}
\ead{constantinoantonio.garcia@usc.es}
\author[ceu]{Abraham~Otero}
\ead{abraham.otero@usc.es}
\author[ourense]{Xos\'{e}~Vila}
\ead{anton@uvigo.es}
\author[citius]{David~G.~M\'{a}rquez}
\ead{david.gonzalez.marquez@usc.es}
\cortext[cor1]{Corresponding author: Tel.: +34 8818 16391. Fax: +34 8818 16405.}
\address[citius]{Centro Singular de Investigaci\'{o}n en Tecnolox\'{i}as da Informaci\'{o}n (CITIUS), University of Santiago de Compostela, 15782, Santiago de Compostela, Spain.}
\address[ceu]{Department of Information and Communications Systems Engineering, University San Pablo CEU, 28668, Madrid, Spain.}
\address[ourense]{Department of Computer Science, University of Vigo, Campus As Lagoas s/n, 32004, Ourense, Spain.}

\begin{abstract}
One of the most promising non-invasive markers of the activity of the autonomic nervous system is Heart Rate Variability (HRV). HRV analysis toolkits often provide spectral analysis techniques using the Fourier transform,  which assumes that
the heart rate series is stationary. To overcome this issue, the Short Time Fourier Transform is often used (STFT). However, the wavelet transform is thought to be  a more suitable tool for analyzing non-stationary signals than the STFT. Given the lack of support for wavelet-based analysis in HRV toolkits, such analysis must be implemented by the researcher. This has made this technique underutilized.

This paper presents a new algorithm to perform HRV power spectrum analysis based on the Maximal Overlap Discrete Wavelet Packet Transform (MODWPT). The algorithm calculates the power in any spectral band with a given tolerance for the band's boundaries. The MODWPT decomposition tree is pruned to avoid calculating  unnecessary wavelet coefficients, thereby optimizing execution time. The center of energy shift correction is applied to achieve optimum alignment of the wavelet coefficients. This  algorithm has been implemented in RHRV, an open-source package for HRV analysis. To the best of our knowledge, RHRV is the first HRV toolkit with support for wavelet-based spectral analysis.

\end{abstract}


\end{frontmatter}


\section{Introduction}

Heart Rate Variability (HRV) refers to the variation over time of the intervals between consecutive heartbeats. 
Since the heart rhythm is modulated by the autonomic nervous system (ANS), HRV is thought
to reflect the activity of  the sympathetic and parasympathetic branches of
the ANS. The continuous modulation of the ANS results in continuous variations in heart rate.
One of the most powerful HRV analysis techniques is based on the spectral analysis of the time series
obtained from the  distances between each pair of consecutive heartbeats. The HRV power spectrum is a useful tool as a predictor of multiple pathologies \cite{malik1995}, \cite{jovic2012evaluating}.

Akselrod et al. \cite{akselrod1981} described three components in the HRV power spectrum
with physiological relevance: the very low frequency (VLF) component (frequencies below 0.03 Hz),
which is modulated by the renin-angiotensin system; 
the low frequency (LF) component (0.03-0.15 Hz), which is thought to be of both sympathetic and
parasympathetic nature; and the high frequency (HF) component
(0.18-0.4 Hz), which is related to the parasympathetic system.
At present, there is no absolute consensus on the precise limits of the
boundaries of these three bands. In the literature we can find authors who use slightly different  bands' boundaries \cite{lewis2007influence}.

There exist several HRV spectral analysis techniques. These techniques may be classified 
as nonparametric and parametric \cite{taskforce1996}. The main advantage
of the nonparametric methods is the simplicity and speed of the algorithm used (The Fast Fourier Transform). The main advantage
of the parametric methods is that they give smoother spectral components. However, parametric methods present problems regarding to correct model order selection. 
Although these techniques are widely used, they have no temporal resolution.  This severely limits their ability 
to analyze non-stationary signals and transient phenomena. To alleviate this limitation temporal windows are often used, so that small segments of the whole signal are analyzed. Among these techniques we may highlight the Short
 Time Fourier Transform (STFT) \cite{vila1997}. However, time-frequency resolution of the STFT depends on the spread of the window used. Thus, the STFT has fixed time-frequency resolution: high frequency resolution implies poor time resolution and vice versa.
  Conversely to Fourier,
the wavelet transform performs time-frequency analysis and it is recognized as
 a powerful tool  to study non-stationary signals \cite{mallatMR}.
 
HRV analysis toolkits such as Kubios HRV \cite%
{kubios} or aHRV
\cite{ahrv} only enable HRV spectral analysis based on the Fourier
transform or parametric methods.
To the best of our knowledge, the only option for using the wavelet transform in HRV analysis
is to manually implement the algorithms, probably with the support of some general wavelet library. This is tedious, and prone
to error. Although some researchers have done this \cite{addison2005wavelet}, \cite{hossen2007wavelet}, many more (especially those with a medical background) choose to use Fourier-based tools, even
when they know that the signal being analyzed is non-stationary. A query in the PubMed database with the terms ``heart rate variability Fourier transform" returns 660 articles, while a query with the terms ``heart rate variability wavelet transform" only returns 145 articles. The lack of tools for carrying out HRV analysis using the wavelet transform has made this potentially superior analysis technique underutilized in comparison with the Fourier transform.

In this paper we present an algorithm to perform HRV power spectrum analysis based on the Maximal Overlap Discrete Wavelet Packet Transform (MODWPT). The algorithm calculates the spectrogram in any frequency band, allowing a certain tolerance for the position of the band's boundaries. The algorithm has been validated over simulated and real RR series. Its capability for identifying  fast changes in the RR series' spectral components has been compared with the STFT and a windowed version of the Burg method, showing that these techniques miss some transient changes that are successfully identified by the wavelet transform. We have implemented the algorithm in RHRV, an open-source package for HRV analysis publicly available on the Internet. A previous version of this algorithm was published in \cite{primero}.

Section \ref{sec:waveletReview} starts with a brief review of the wavelet transform, with particular attention to the MODWPT, and then introduces our algorithm to perform HRV power spectrum analysis. Section \ref{sec:swDes} provides a short description of the implementation of the algorithm in the RHRV package. Section \ref{sec:Results} presents a comparison between our algorithm, the STFT and the windowed Burg method over simulated and real RR series. Finally, the results of this paper are discussed and some conclusions are given.
\section{Material and methods\label{sec:waveletReview}}
A brief review of some important wavelet concepts for our algorithm is now given.
A wavelet is a small wave $\psi(t)$ (oscillating function) that is well concentrated in time. This function must
have unitary norm $\|\psi\|=1$ and verify the so-called admissibility condition: $\int_{-\infty}^{\infty}\psi (t)dt=0.$
$\psi (t)$ can be translated and dilated in time, yielding a set of wavelet functions:

\begin{equation}
\psi _{u,s}(t)=\frac{1}{\sqrt{s}}\psi \left( \frac{t-u}{s}\right),
\end{equation}%
where $s>0$ is a dilation factor, and $u$ is a real number
representing the translations. As $\psi$ generates all $\psi_{u,s}$
functions, it is called mother wavelet.

A continuous wavelet transform measures the time-frequency variations of a signal $f$ by correlating it with $\psi_{u,s}$

\begin{equation}
Wf(u,s)=\int_{-\infty}^{\infty}f(t)\psi^*_{u,s} (t)dt.
\end{equation}

In order to make the wavelet transform implementable on a computer, both dilation and translation factors
must be discretized. This can be achieved as follows:
\begin{equation}
\left\{\psi _{j,n}=\frac{1}{\sqrt{2^{j}}}\psi \left( \frac{t-2^{j}n}{2^{j}}%
\right) \right\}_{j,n\;\in\; \mathbb{Z}.}
\end{equation}%

This family is an orthonormal basis of $\mathbf{L^{2}(\mathbb{R})}$. Orthogonal wavelets dilated by $2^j$ can be used
to study signal variations at the resolution $2^{-j}$. Thus, these families of wavelets originate a
multiresolution signal analysis. Multiresolution analysis projects signals at various resolution spaces  $\mathbf{V}_{j}$. 
Each  $\mathbf{V}_{j}$ space contains all possible approximations at the resolution $2^{-j}$. Thus, each decomposition level
increases the spectral resolution of the decomposition, at the expense of losing
temporal resolution.
Let $\{\mathbf{V}%
_{j}\}_{j\in
\mathbb{Z}
}$ be a multiresolution approximation verifying $%
\mathbf{V}_{j+1}\subset \mathbf{V}_{j}\;\forall j\in
\mathbb{Z}
$ and let $\mathbf{W}_{j}$ be the orthogonal complement of
$\mathbf{V}_{j}$ in $\mathbf{V}_{j-1}$: $\mathbf{V}_{j-1}=\mathbf{V}_{j}\oplus
\mathbf{W}_{j}$. According to \cite{mallatWTSP}, the families
\begin{equation}
\left\{\phi _{j,n}=\frac{1}{\sqrt{2^{j}}}\phi \left( \frac{t-2^{j}n}{2^{j}}%
\right) \right\}_{n\in \mathbb{Z}}
\qquad \text{ and } \qquad
\left\{\psi _{j,n}=\frac{1}{\sqrt{2^{j}}}\psi \left( \frac{t-2^{j}n}{2^{j}}%
\right) \right\}_{n\in \mathbb{Z}}
\end{equation}%
are an orthonormal basis for $\mathbf{V}_{j}$ and $\mathbf{W}_{j}$,
respectively, for all $j\in
\mathbb{Z}
$. $\psi _{j,n}$ are the wavelet functions and $\phi _{j,n}$\ are
the scaling functions.

Thus, we can approximate any function $f$
$\epsilon\;\mathbf{L^{2}(\mathbb{R})}$ at resolution $2^{-j}$ as
\begin{equation}
P_{\mathbf{V}_{j}}f=\sum_{n=-\infty }^{\infty }\langle f,\phi
_{j,n}\rangle \phi _{j,n}=\sum_{n=-\infty }^{\infty }a_{j}[n]\phi
_{j,n}
\label{Eq:resolutionProjection}
\end{equation}%
and the orthogonal projection of $f$ onto detail space
$\mathbf{W}_{j}$ is:
\begin{equation}
P_{\mathbf{W}_{j}}f=\sum_{n=-\infty }^{\infty }\langle f,\psi
_{j,n}\rangle \psi _{j,n}=\sum_{n=-\infty }^{\infty }d_{j}[n]\psi
_{j,n}.
\label{Eq:detailProjection}
\end{equation}%
where $a_{j}[n]$ and $d_{j}[n]$ are called the approximation and
detail coefficients, respectively.

Mallat proved \cite{mallatMR} that both approximation and detail
coefficients may be calculated using a filter bank. Let $h[n]$ and $g[n]$ be the
FIR filters that will be used to compute the approximation and
detail coefficients, respectively. It has been proven
\cite{mallatWTSP} that the filter $h[n]=\langle \frac{1}{\sqrt{2}}\phi\left(\frac{t}{2}\right),
\phi\left(t-n\right)\rangle $ and that $g[n]=\langle \frac{1}{\sqrt{2}}\psi\left(\frac{t}{2}\right),
\phi\left(t-n\right)\rangle $. $g[n]$ and $h[n]$ can be regarded as an approximation to a high-pass
filter (the wavelet filter) and to a
low-pass filter (the scaling filter), respectively. By applying recursively over the approximation
coefficients the same filtering operation followed by sub-sampling
by two, we obtain the multiresolution expression of $f$. This
algorithm, known as the pyramid algorithm, is the most efficient way
of computing the Fast Orthogonal Wavelet Transform (FOWT) \cite{mallatWTSP}.

\subsection{MODWPT\label{sec:mo}}
Given that the filtering operation is only applied over the approximation
coefficients, the FOWT only provides information on a limited set of
frequency bands which need not be the ones used in the HRV analysis. A more suitable 
wavelet transform is needed for our algorithm: the wavelet packet decomposition (WPD).
 Instead of dividing only the approximation coefficients $a_j[n]$, both detail and approximation coefficients  are decomposed 
 successively by applying high pass and
low pass filters to each set of coefficients.

Among the WPD transforms we have chosen the MODWPT \cite{percival2006} because it is less
sensitive to the starting point of the time series and it is applicable
to non dyadic sequences. Furthermore, the MODWPT avoids the sub-sampling step,
and therefore it has the same number of wavelet coefficients in
every decomposition level. This simplifies computations involving
different decomposition levels.

\begin{figure}[ht]
\begin{center}
\includegraphics[width=3.5in]{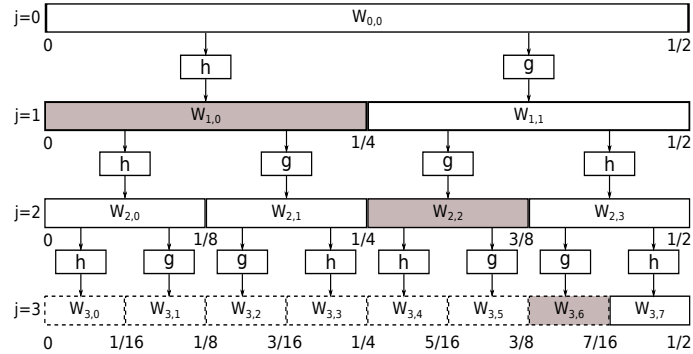}
\end{center}
\caption{MODWPT decomposition tree with the nodes selected to cover the band $[0,7/16]$ Hz. $\mathbf{W}_{0,0}$ represents
the original signal, $f(t).$} \label{modwpt}
\end{figure}

The $j^{th}$  level of the MODWPT decomposes the frequency interval $%
[0,f_{s}/2] $, where $f_{s}$ is the sampling frequency of the
original signal $f$, into $2^{j}$ equal width intervals (see Fig.
\ref{modwpt}). Thus, the $n^{th}$ node (beginning at zero) 
in the $j^{th}$ level of the decomposition tree, the $(j,n)$ node, is
associated with the frequency interval $\frac{f_{s}}{2^{j+1}}\left[
n,n+1\right] $. Each node will have $N$ wavelet
coefficients associated, $N$
being the length of the sampled signal $f$. The $N$-dimensional vector $%
\mathbf{W}_{j,n}$ will denote the $N$ wavelet coefficients
associated with the $(j,n)$ node. 

MODWPT coefficients fulfill that:%
\begin{equation}
\Vert f\Vert ^{2}=\sum_{n=0}^{2^{j}-1}\Vert \mathbf{W}_{j,n}\Vert
^{2}\quad \forall j.
\end{equation}

Therefore, given a frequency band $\left[ f_{1},f_{2}\right] =\frac{f_{s}}{%
2^{j+1}}\left[ k,k^{\prime }+1\right] $, $f_{s}$ being the sampling
frequency and $k$, $k^{\prime }$ and $j$ integers,
the spectral power in $\left[ f_{1},f_{2}\right] $, $P(\left[
f_{1},f_{2}\right] )$, can be calculated from the
appropriate MODWPT coefficients. We just need to find the nodes $(j,k)$ and $%
(j,k^{\prime })$ and compute the spectral power in the band
$\left[
f_{1},f_{2}\right] $ as:%
\begin{equation}
P(\left[ f_{1},f_{2}\right] )=\sum_{n=k}^{k^{\prime }}\Vert \mathbf{W}%
_{j,n}\Vert ^{2}.  \label{Eq:PowerInBand}
\end{equation}

\subsection{Finding a proper cover for a set of frequency bands\label{sec:cover}}

Equation (\ref{Eq:PowerInBand}) can only be applied to bands that can
be written as $\frac{f_{s}}{2^{j+1}}\left[ k,k^{\prime }+1\right] $,
$f_{s}$ being the sampling frequency and $k$, $k^{\prime }$ and $j$
integers. In an HRV spectral analysis the user
can be interested in bands that cannot be written this way. This
forces us to permit a certain error when we try to cover the bands with coefficients obtained from the MODWPT
decomposition. Let $[f_{l},f_{u}]$ be the band, and let $\epsilon _{l}$ and $\epsilon _{u}$ be the
maximum errors allowed for the beginning and the ending of the band,
respectively. We need
to find a node $(j,n)$ whose lower frequency corresponds roughly to
$f_{l}$ with the tolerance allowed by $\epsilon _{l}$, i.e.:
\begin{equation}
\left|f_{l}-\frac{f_{s}}{2^{j+1}}n\right|\leq \epsilon _{l}.  \label{c2l}
\end{equation}%

Analogously, we also need to find a node $(j^{\prime
},n^{\prime })$ whose upper frequency corresponds roughly to $f_{u}$
with the tolerance allowed by $\epsilon _{u}$:
\begin{equation}
\left|f_{u}-\frac{f_{s}}{2^{j^{\prime }+1}}(n^{\prime }+1)\right|\leq \epsilon
_{u}. \label{c2u}
\end{equation}%

We shall refer to (\ref{c2l}) and (\ref{c2u})  as \textit{the cover conditions}.

The level $j$ of the decomposition tree in which the node $(j,n)$
that fulfills (\ref{c2l}) is found needs not
be the same as the level $j^{\prime }$ in which the node $(j^{\prime
},n^{\prime })$ that
fulfills (\ref{c2u}) is found. But, (\ref%
{Eq:PowerInBand}) requires that $j=j^{\prime }$. To avoid this
problem we may think that, after the nodes $(j,n)$ and
$(j^{\prime },n^{\prime } )$ have been found, we could 
descend the node that is at the higher level, decomposing it
to the level of the other node \cite{primero}. However, when descending to low levels in
the frequency decomposition tree, frequency resolution increases and
temporal resolution decreases because 
of the Gabor-Heisenberg uncertainty principle for signals: 
$\Delta t\Delta f\geq 1/2$ \cite{gabor1946}. 
Furthermore, wavelet coefficients suffer a circular time shift as a result
of the use of wavelet filters. As we descend to lower levels, the circular shift will be more
pronounced as we perform more convolutions.
Therefore, to obtain a good temporal
resolution, we should avoid descending to deep levels of the tree.	

Let's suppose that,
when looking for a cover for the band of interest $\left[ f_1,f_2\right]$, the nodes $(j,n)$ and $(j^{\prime
},n^{\prime })$ are selected. These nodes must fulfill the 
cover conditions (\ref{c2l}) and (\ref{c2u}) with $f_{l}=f_{1}$ and $f_{u}= f_{2}$. Generally, the nodes $(j,n)$ and $(j^{\prime
},n^{\prime })$  are at different levels (i.e., $j \neq j^{\prime}$). Furthermore,
 these nodes cover the band
$\frac{f_{s}}{2^{j+1}}\left[ n,n+1 \right] 
\cup 
\frac{f_{s}}{2^{j^{\prime }+1}} \left[ n^{\prime },n^{\prime }+1 \right]$, but the band
 $f_{s}\left[ \frac{n+1}{2^{j+1}},\frac{n^{\prime}}{2^{j^{\prime}+1}}
  \right]=\left[ f^{\prime}_1,f^{\prime}_2\right]$ remains uncovered (see nodes $(1,0)$ and $(3,6)$ of Fig. \ref{modwpt}).
The problem has been reduced to finding a cover for the band $\left[ f^{\prime}_1,f^{\prime}_2\right]$. Thus, new nodes will be selected fulfilling 
\textit{the cover conditions} (\ref{c2l}) and (\ref{c2u}), with $f_l=f^{\prime}_1$ and $f_u=f^{\prime}_2$.
The complete cover of $\left[ f_1,f_2\right]$ can be achieved applying
this technique recursively. The set of nodes resulting from this
cover will be referred to as  $\Gamma$.

There may be overlap between the bands associated with the nodes $(j,n)$ and $(j^{\prime },n^{\prime })$ (the original nodes fulfilling the \textit{cover conditions}). This
overlap will occur if a node is the parent of the other node or if both nodes are equal. If $(j,n)=(j^{\prime 
},n^{\prime })$, the node covers all the band and the search has ended. If the node $(j^{\prime },n^{\prime })$ is a child of the node $(j,n)$, we shall replace the 
node $(j,n)$ with its (unique) child verifying (\ref{c2l}): the node $(j+1,2n)$. Thus, the algorithm continues
with the nodes $(j+1,2n)$ and $(j^{\prime },n^{\prime })$. The process shall be repeated until there is no overlap between the two nodes fulfilling the \textit{cover condition}. Similarly, if the node $(j,n)$
is a child of the node $(j^{\prime },n^{\prime })$, we replace the node $(j^{\prime },n^{\prime })$  with the node $(j^{\prime }+1,2n^{\prime }+1)$. 

Figure \ref{modwpt} illustrates the cover process
for the band $\left[0,7/16\right]$ Hz with $f_s = 1$ Hz and 
$\epsilon_u =\epsilon_l = 0.01$ using the ``nodes at the same
level" criteria used in \cite{primero} (dashed nodes) and the  ``nodes at different
levels" criteria  proposed here (dark nodes).

There exist multiple covers of any band $\left[ f_{l},f_{u}\right]$. Our algorithm builds a cover using the nodes of the higher levels of the MODWPT tree. This selection criteria leads to a cover that minimizes temporal shift and maximizes temporal resolution. Furthermore, this cover also \textit{minimizes} the number of nodes. 

Equation (\ref{Eq:PowerInBand}) cannot
be applied to this decomposition, because the decomposition uses nodes
from different levels. However, because 
of the wavelet coefficients property

\begin{equation}\Vert \mathbf{W}_{j,n}\Vert^2 =\Vert \mathbf{W}_{j+1,2n}\Vert^2+\Vert \mathbf{W}_{j+1,2n+1}\Vert^2,
\end{equation}
we may write (\ref{Eq:PowerInBand}) as
\begin{equation}
P(\left[ f_{1},f_{2}\right] )\approx\sum_{n=k}^{k^{\prime }}\Vert \mathbf{W}%
_{j,n}\Vert ^{2}=\sum_{\mathbf{W}_{j,n}\;\epsilon\;\Gamma}\Vert \mathbf{W}%
_{j,n}\Vert ^{2}.  \label{Eq:NewPowerInBand}
\end{equation}
 
\begin{figure}[ht]
\begin{center}
\includegraphics[width=3.5in]{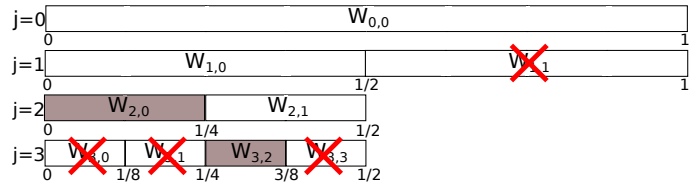}
\end{center}
\caption{Prune procedure using the PMODWPT. The crosses indicates which nodes have been pruned.} \label{prune}
\end{figure}

In practical applications, the complete MODWPT decomposition tree will be rarely needed.
In order to improve the execution time of wavelet analysis we have developed a pruning algorithm that we call Pruned MODWPT (PMODWPT). The PMODWPT, instead of expanding
every node of the decomposition tree will only calculate those children required to
cover a frequency band. Figure \ref{prune} illustrates the pruning performed when calculating the power of the band
$\left[0,3/8\right]$ Hz being $f_s = 2$ Hz. 

\subsection{Shift correction\label{sec:shift}}
\begin{figure}[ht]
\begin{center}
\includegraphics[angle=0,width=0.6\textwidth]{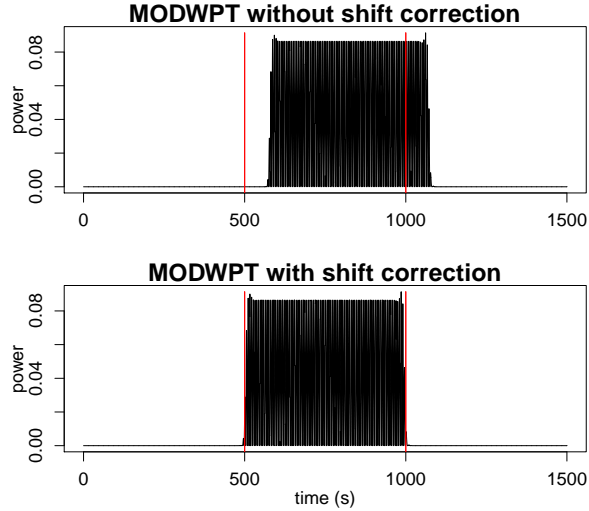}
\caption{Time shift using MODWPT without (top) and with (bottom) ``Center of Energy" correction. The vertical lines indicate where spectral power should be.} \label{shifts}
\end{center}
\end{figure}

As a consequence of the frequency-dependent phase response of the wavelet filters, it is necessary to keep track of which wavelet coefficients contain the energy contribution of a given portion of the signal being analyzed (see Fig. \ref{shifts}). That
is, the wavelet coefficients must be corrected so that they can be accurately
aligned with the original time series. 

Hess-Nielsen and Wickerhauser proposed in \cite{hessWicker} a technique
to compute shift corrections exact for linear phase
filters, and showed how to estimate the perturbation that a deviation from
linear phase produces. This suggests that the method will produce better 
results with filters whose phase does not deviate much from a linear phase filter. 
Thus, this algorithm can be used to compute approximate shift corrections for a wide range 
of wavelets, including least asymmetric and extremal phase Daubechies, Symmlet or Coiflet wavelets.

Time shifts proposed in \cite{hessWicker} are based on the notion
of the ``center of energy" of a filter. Let $a_l$ be a filter
of length $l$. Its center of energy,  
$E\left\lbrace a_l\right\rbrace$ ,
is given by:

\begin{equation}
E\left\lbrace a_l\right\rbrace = \frac{\underset{n=0}{\overset{l-1 }{\sum}} n\cdot a_l[n]^2}{\underset{n=0}{\overset{l-1 }{\sum}} a_l[n]^2}
\end{equation}

The time shift needed for the node $(j,n)$ of the MODWPT is always an advance
(shift to the left) of $| p_{j,n}|$ units that is given  by \cite{hessWicker}
\begin{equation}
|p_{j,n}|=C_{j,n}(E\left\lbrace g\right\rbrace-E\left\lbrace h\right\rbrace) + (2^j-1)E\left\lbrace h\right\rbrace, 
\label{shiftEq}
\end{equation}
where $C_{j,n}$ is the bit-reversal of the binary gray code which encodes each of the MODWPT tree nodes \cite{hessWicker}.

\subsection{Wavelet-based HRV analysis}
Listing \ref{lst:wsa} shows the pseudocode of our wavelet-based HRV analysis algorithm.
The function \textit{waveletAnalysis} takes as input parameters the interpolated and filtered RR series, the boundaries of the frequency bands, the wavelet to be  used in the analysis, the sampling frequency of the RR series and the tolerance allowed when covering the frequency bands to be analyzed. The algorithm starts by finding the covers for each of the bands in which the user wants to calculate the spectral power. The \textit{calculateCover} function implements the recursive covering algorithm presented in section \ref{sec:cover}. The \textit{getNode} function selects the initial nodes fulfilling the cover conditions given by (\ref{c2l}) and (\ref{c2u}). It expands the node whose frequency interval contains the boundary frequency ($f$). If none of its children verify the cover condition for $f$, the child containing $f$ is selected and the process is repeated. The \textit{fillGapsBetweenNodes} function completes the cover given a pair of initial nodes: $(j,n)$ and $(k,m)$ that fulfill  (\ref{c2l}) and (\ref{c2u}), respectively. This function distinguishes the four cases  discussed in section \ref{sec:cover}: $(j,n)=(k,m)$; $(k,m)$ is a child of the $(j,n)$ node; $(j,n)$ is a child of $(k,m)$ or the general case where there is an uncovered band. 

Once the cover for the bands has been found, the wavelet coefficients are computed using the PMODWPT. Note that we use a single decomposition tree to calculate the power in all the bands. The shift correction described in section \ref{sec:shift} is computed using \textit{shiftCorrection}. The target nodes are specified to avoid applying corrections to nodes that will not be used in subsequent calculations. Since the wavelet and scaling filters are involved in (\ref{shiftEq}), the wavelet used to perform the analysis must also be specified in the \textit{shiftCorrection} function. Finally, the spectrogram of each band is computed with $\sum_{\mathbf{W}_{j,n}\;\epsilon\;\Gamma}| \mathbf{W}_{j,n}|^2$ by the \textit{getSpectrogram} function.

\lstset{linewidth=0.9\textwidth}
\begin{lstlisting}[mathescape,caption={Pseudocode of our wavelet-based HRV analysis algorithm.},label={lst:wsa}]
waveletAnalysis($rr$, $VLFmin$, $VLFmax$, $LFmin$, $LFmax$, $HFmin$, $HFmax$,$wavelet$, $f_s$,$bandtolerance$){ 
  $VLFnodes$=calculateCover($[VLFmin,VLFmax]$,$f_s$,$bandtolerance$)
  $LFnodes$=calculateCover($[LFmin,LFmax]$,$f_s$,$bandtolerance$)
  $HFnodes$=calculateCover($[HFmin,HFmax]$,$f_s$,$bandtolerance$) 
  $wrr$=PMODWPT($rr$,$wavelet$,$VLFnodes\cup LFnodes\cup HFnodes$)
  $wrr$ = shiftCorrection($wrr$,$wavelet$,$VLFnodes\cup LFnodes\cup HFnodes$)
  $VLFpower$=getSpectrogram($wrr$,$VLFnodes$)
  $LFpower$=getSpectrogram($wrr$,$LFnodes$)
  $HFpower$=getSpectrogram($wrr$,$HFnodes$)
  return($VLFpower$,$LFpower$,$HFpower$)
}

calculateCover([$f_l$,$f_u$],$f_s$,$error$){
   if ($f_u$==$f_l$ ) return($\left\{\right\}$)
   $W_{j,n}$=getNode($f_l$,$f_s$,$error$)
   $W_{k,m}$=getNode($f_u$,$f_s$,$error$)
   return(fillGapsBetweenNodes($W_{j,n}$,$W_{k,m}$,$f_s$,$error$))
}

getNode($f$,$f_s$,$error$,$type$){   
    $i$=$1$
    $nodeToExpand=0$
    while($TRUE$){
          $b_j$=$nodeToExpand$
          $\Delta$ = $\frac{f_s}{2^{i+1}}$
          for ($j$ in $(2\cdot b_j):(2\cdot b_j+1)$){
            $interval$= $[j\Delta,(j+1)\Delta]$
            if ($f$ $\in$ $interval$) $nodeToExpand$=$j$
            if ($interval$ verifies band cover condition for $f$ and $error$) return ($W_{i,j}$)
          }
          $i=i+1$
    }
}
 
fillGapsBetweenNodes($W_{j,n}$,$W_{k,m}$,$f_s$,$error$){
   if($W_{j,n}$==$W_{k,m}$) return($W_{j,n}$)
   if($W_{k,m}$ is Child of $W_{j,n}$) return(fillGapsBetweenNodes($W_{j+1,2n}$,$W_{k,m}$,$f_s$,$error$))
   if($W_{j,n}$ is Child of $W_{k,m}$) return(fillGapsBetweenNodes($W_{j,n}$,$W_{k+1,2m+1}$,$f_s$,$error$))
   $mediumNodes$=calculateCover($f_s\left[ \frac{n+1}{2^{j+1}},\frac{m}{2^{k+1}} \right]$,$f_s$,$error$)
   return($\left\{W_{j,n},W_{k,m}\right\}\cup\;mediumNodes$)
}

getSpectrogram($wrr$,$nodes$){
  $S_{\left\{\mathbf{x}\right\}}[m]=0$
  for ($W_{j,n}$ in $nodes$){
    $d_{j,n}[m]$=getWaveletCoefficients($wrr$, $W_{j,n}$) 
    $S_{\left\{\mathbf{x}\right\}}[m]=S_{\left\{\mathbf{x}\right\}}[m]+|d_{j,n}[m]| ^{2}$
  }
  return ($S_{\left\{\mathbf{x}\right\}}[m]$)
}

\end{lstlisting}

\subsection{Implementation of the algorithms in RHRV\label{sec:swDes}}
RHRV is an open-source package for the R environment
for statistical computing that comprise a complete set of
tools for HRV analysis. Further details about RHRV package may be found in \cite{vilaRHRV}. Here we shall only describe the RHRV functionality which is related to the algorithm presented in this paper. 

 RHRV imports data files containing heartbeat positions. Supported formats include ASCII (\textit{LoadBeatAscii} function), EDF (\textit{LoadBeatEDFPlus}), Polar (\textit{LoadBeatPolar}), Suunto (\textit{LoadBeatSuunto}) and WFDB data files (\textit{LoadBeatWFDB}). To compute the instantaneous heart rate series \textit{BuildNIHR} can be used. A filtering operation can be carried out in order to eliminate outliers or spurious points present in the HR time series with
unacceptable physiological values (\textit{FilterNIHR}). A uniformly sampled heart rate signal (with equally spaced values) is obtained using
\textit{InterpolateNIHR}.

Spectral power HRV analysis is performed with the
\textit{CalculatePowerBand} function.  This function computes
the spectrogram of the heart rate series in the  ULF,
VLF, LF and HF frequency bands. Boundaries of the bands may be chosen 
by the user. If boundaries are not specified, default values 
are used: ULF, $\left[ 0,0.03\right] $ Hz;
VLF, $\left[ 0.03,0.05\right] $ Hz;  LF, $\left[ 0.05,0.15\right] $ Hz; HF, $%
\left[ 0.15,0.4\right] $ Hz. 
Until now, \textit{CalculatePowerBand} used the STFT to compute 
the spectral power (window size, window shift and zero padding may  be specified by the user). The wavelet analysis algorithm
presented in this paper was included in this function  in such a way that we maintain backward compatibility. Thus, both Fourier and wavelet analysis may be used with the \textit{CalculatePowerBand} function. Type of analysis can be selected by the user by specifying the \textit{type}
parameter (``fourier" or ``wavelet"). 

When using wavelet analysis, in addition to the frequency bands, an error for the boundaries (default value is 0.1 in absolute terms) and a mother wavelet can be specified by the user. Some of the most used wavelets are available: ``haar", extremal phase (``d4", ``d6", ``d8" and ``d16")
and the least asymmetric (``la8", ``la16" and ``la20") Daubechies,
and the best localized (``bl14" and ``bl20")
among others. Default value is ``d4". Listing \ref{lst:basicexample} shows all the RHRV code required to perform a typical wavelet-based spectral analysis.

\lstset{linewidth=0.9\textwidth}
\begin{lstlisting}[language=R,caption={HRV wavelet-based analysis in
RHRV.},label={lst:basicexample}]
md = CreateHRVData( )
md = LoadBeatAscii(md, "BeatPositions.beats")
md = BuildNIHR(md)
md = FilterNIHR(md)
md = InterpolateNIHR(md, freqhr = 4)
md = CreateFreqAnalysis(md)
md = CalculatePowerBand(md, indexFreqAnalysis=1, ULFmin = 0, ULFmax=0.03, VLFmin = 0.03,VLFmax = 0.05, LFmin = 0.05, LFmax = 0.15, HFmin= 0.15, HFmax =0.4, type="wavelet", wavelet="d4", bandtolerance=0.01)
PlotPowerBand (md, indexFreqAnalysis=1)
\end{lstlisting}

\section{Results\label{sec:Results}}
\subsection{Temporal resolution}
In order to compare the temporal resolution of the HRV analysis techniques based on our algorithm with those
based on the Fourier transform (the STFT) and parametric estimation (the windowed Burg method), simulated RR series will be used. Simulated signals  are used
instead of real signals because when using real signals, it is difficult to know which is the ``correct" result. Therefore, we use simulated signals to know exactly what spectral components they have at each instant. 

The Integral Pulse Frequency Modulation (IPFM) model \cite{berger1986}, \cite{hyndman}
is a widely accepted technique used to generate RR series. The IPFM model simulates the sino-atrial node (SA)
modulation by using a modulating signal $m(t)$ and  the SA function 
as trigger of the cardiac contraction by using a threshold $\hat{T}$. We shall use a signal with several fast (every 16 seconds) spectral changes as modulating signal:

\begin{equation*}
\footnotesize
m(t)=\left\{
\begin{tabular}{l}
$0.3\sin {(2\pi \cdot 0.09375\cdot t)}\;\;0\leq t<16 \;s$ \\
$0.3\sin {(2\pi \cdot 0.03125\cdot t)}\;\;16\leq t<32 \;s$ \\
$0.3\sin {(2\pi \cdot 0.09375\cdot t)}\;\;32\leq t<48 \;s$ \\
$0.3\sin {(2\pi \cdot 0.03125\cdot t)}\;\;48\leq t<64 \;s$ \\
$0.3\sin {(2\pi \cdot 0.09375\cdot t)}\;\;64\leq t<80 \;s$ \\
\end{tabular}%
\right.
\end{equation*}

The idea is to test the time-frequency transforms in a non-stationary scenario. To obtain a more realistic simulation, white noise was added to the $m(t)$ signal.

\begin{figure}[H]
\begin{center}
\begin{tabular}{ll}
\subfigure[Spectrogram using STFT HRV analysis.]{
 	\includegraphics[angle=0,width=2.0in]{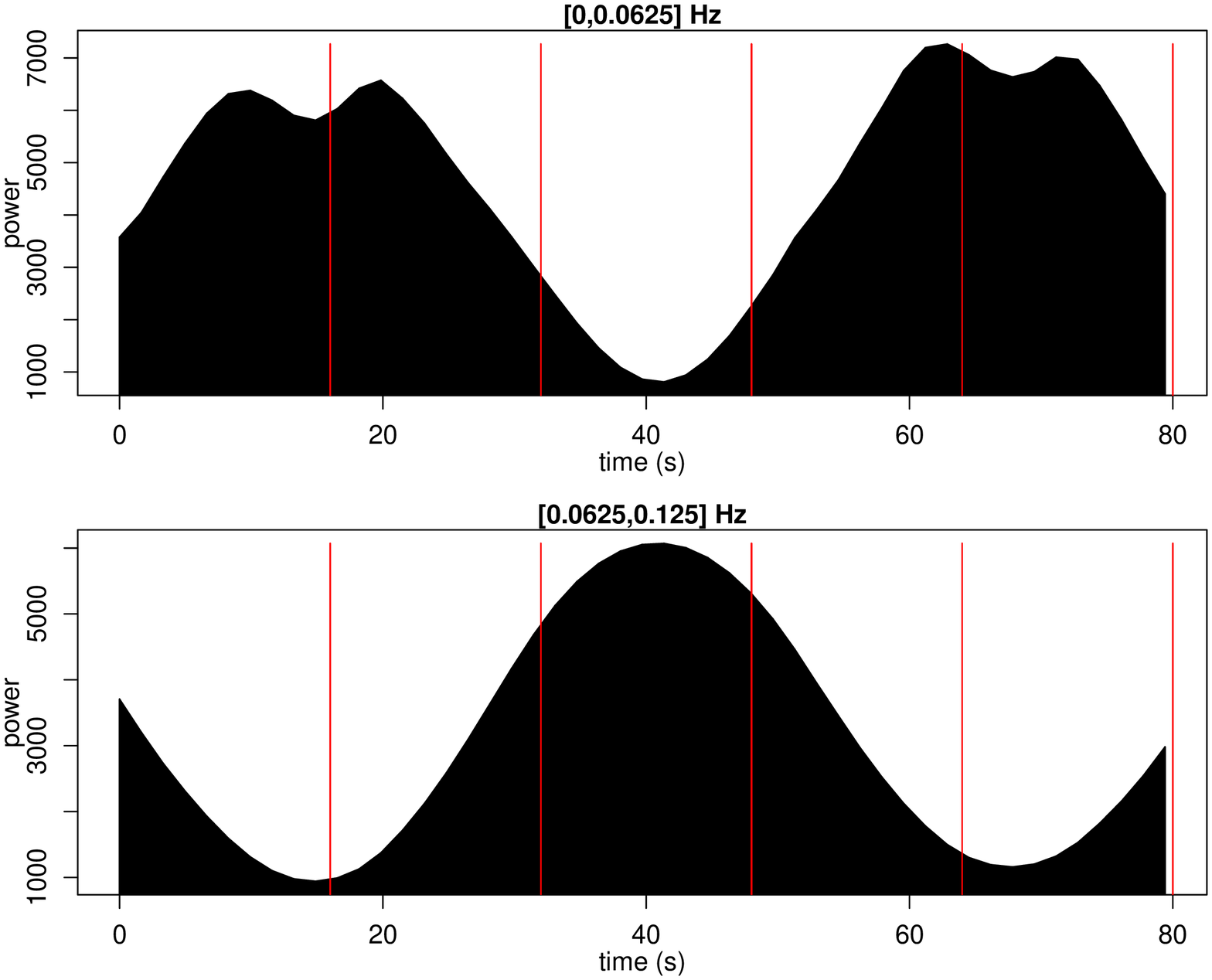}
 	\label{fourierAnNS}	 
}
\subfigure[Spectrogram using parametric HRV analysis.]{
	\includegraphics[angle=0,width=2.0in]{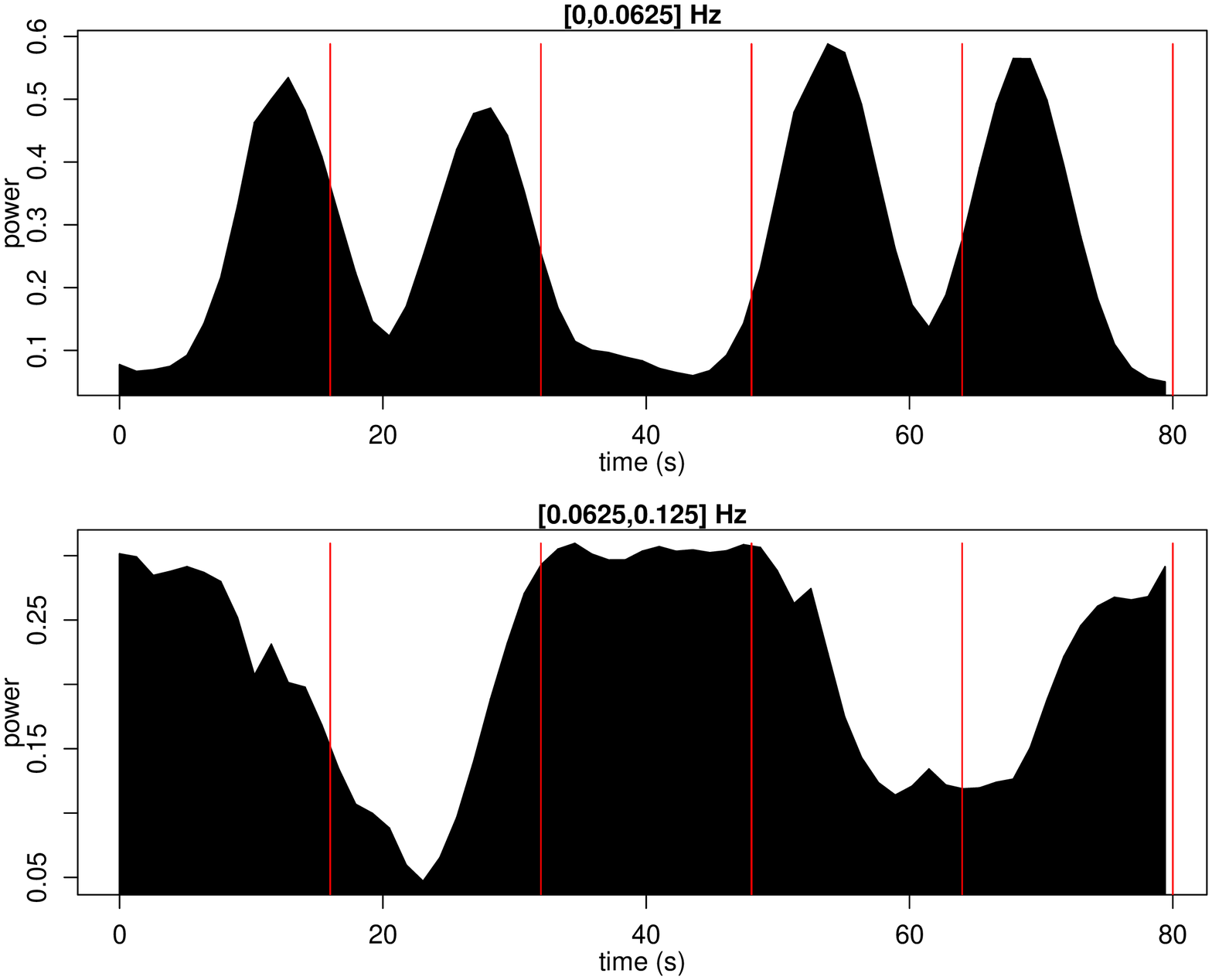}
	 \label{parametricAnNS}	 
}
\end{tabular}
\subfigure[Spectrogram using wavelet HRV analysis.]{
	\includegraphics[angle=0,width=2.25in]{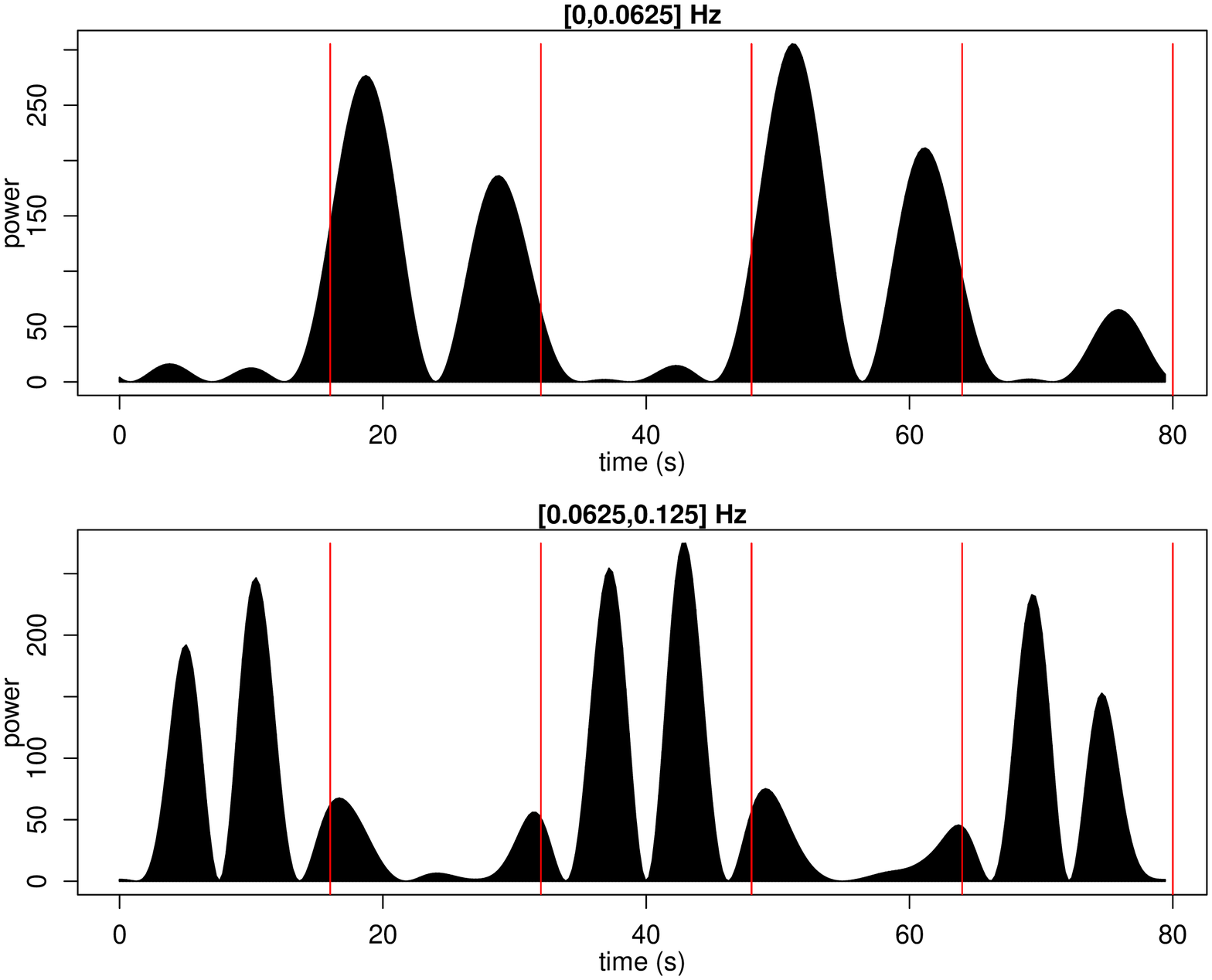}
   \label{waveletAnNS}	 
}

\caption{Spectrogram analysis of the simulated RR series.}
\label{simPower}
\end{center}
\end{figure}

Figure \ref{simPower} shows how the wavelet transform correctly finds most of the spectral power in the first band between 16 s and 32 s, and between 48 s and 64 s; and in the second band between 0 s and 16 s, between 32 s and 48 s and between 64 s and 80s, while the other methods cannot track the spectral changes. The wavelet analysis was performed using a least asymmetric Daubechies filter of width 8. The tolerance was set to 0.01. The selection of the window parameters for the STFT and the Burg method is not trivial. Note that the minimum size of the window for the STFT should be approximately $T\approx 1/0.03125=32\;s$. However, there is a spectral change every 16 seconds. The results shown in Fig. \ref{fourierAnNS} were obtained using  a 30-second window with a 1-second shift. In the Burg method, we may use smaller windows, provided that we take enough points to estimate the model in each segment. The parametric analysis shown in Fig. \ref{parametricAnNS} was performed using  a 16-second window with a 1-second shift. The selected value for the model order was 16 \cite{boardman2002study}. It can be appreciated how the STFT and the Burg method cannot track changes on this signal. Using smaller length analysis windows did not improve these results.

\begin{table}[width=\textwidht,h]
\caption[Power table for the Fourier and wavelet analysis]{Relative power per frequency and per zone for STFT, windowed Burg method and wavelet analysis.} 
\label{tableQuick}
\begin{center}
\subtable[Fourier Analysis]{
\begin{tabular}{cccccc}
\hline
band$\setminus$time(s)& [0,16) & [16,32)& [32,48)& [48,64)& [64,80)\\
\hline 
VLF& 0.209&0.235&0.067&0.217&0.272\\ 
LF&0.128&0.147&0.384&0.224&0.117\\ \hline 
\end{tabular}
\label{tableQuickF}}

\subtable[Parametric Analysis]{
\begin{tabular}{cccccc}
\hline
band$\setminus$time(s)& [0,16) & [16,32)& [32,48)& [48,64)& [64,80)\\
\hline 
VLF&0.180&0.245&0.083&0.259&0.233\\ 
LF&0.228&0.124&0.288&0.167&0.193\\ \hline 
\end{tabular}
\label{tableQuickParametric}}

\subtable[Wavelet analysis]{
\begin{tabular}{cccccc}
\hline
band$\setminus$time(s)& [0,16) & [16,32)& [32,48)& [48,64)& [64,80) \\
\hline 
VLF&0.040&0.400&0.041&0.447&0.072\\ 
LF&0.270&0.071&0.344&0.077&0.238\\ \hline 
\end{tabular}
\label{tableQuickW}}
\end{center}
\end{table}

For each of the two bands, and for each of the five zones with
different spectral components, we calculated the ratio of the power
that the band presents in each zone divided by the overall power of
the band in the five zones. Theoretically, the power in the VLF
band should be distributed among the $2^{nd}$ and $4^{th}$ zones, whereas the power in the LF band
should be distributed among the  $1^{st}$,  $3^{rd}$ and  $5^{th}$ zones. Therefore, the ideal ratios if perfect time-frequency discrimination is obtained are $(0,0.5,0,0.5,0)$ and $(1/3,0,1/3,0,1/3)$. Tables \ref{tableQuickF}, \ref{tableQuickParametric} and \ref{tableQuickW} show the real ratios
for the STFT, the Burg method and wavelet analysis, respectively. We can see that the wavelet ratios are closer to the theoretical values.

\subsection{Computational burden}
\begin{figure}[ht]
\begin{center}

\includegraphics[width=3in]{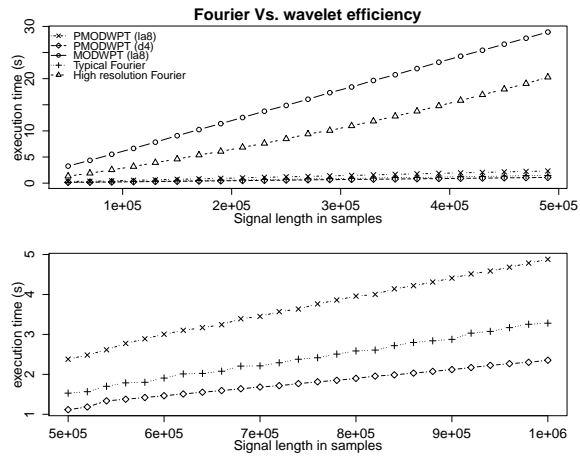}
\end{center}
\caption{Performance of HRV analysis using 
PMODWPT, MODWPT and STFT.} \label{comparativa}
\end{figure}

The top of Fig. \ref{comparativa} compares 
execution time (as a function of the signal size) of HRV analysis algorithms based on MODWPT, PMODWPT (our algorithm)  and the STFT. The input signal to these algorithms was generated randomly.
In order to compare Fourier with wavelet-based analysis, two configurations of the STFT typically used on HRV analysis were selected.
First Fourier analysis 
used a window size and a displacement value of
5 minutes and 30 s, respectively (``Typical Fourier" in Fig. \ref{comparativa}). The second Fourier analysis
used a shorter window in order to achieve a higher temporal resolution. Window size and  displacement took a value
of 30 s and 2.5 s, respectively (``High Resolution Fourier" in
Fig. \ref{comparativa}).
Wavelet analysis was performed using least asymmetric Daubechies of width 8 (``la8") and extremal phase Daubechies of width 4 (``d4") since the efficiency depends on
the filter length. 

PMODWPT and ``Typical Fourier" 
STFT are much more
efficient than MODWPT and ``High Resolution Fourier" STFT when analyzing HRV signals (see the top of Fig. \ref{comparativa}). The bottom
of Fig. \ref{comparativa} shows that the performance of the PMODWPT  is comparable to the ``Typical Fourier" analysis.

\subsection{Validation on real data}
We have tested the algorithms presented in this paper on the recordings of the Apnea ECG database used in the 
PhysioNet/Computers in Cardiology Challenge 2000 \cite{apneaecg}. Obstructive Sleep Apnea-Hypopnea (OSAH) Syndrome is a sleep-breathing disorder
 characterized by the presence of total (apneas) and/or partial (hypopneas) cessations of respiratory airflow while 
 the patient is asleep. There is an interest in developing low-cost OSAH screening tests that can be carried 
out in the patient's home and that can decrease the workload of hospitals' Sleep Units. This is related to the goal of the Computers in Cardiology Challenge 2000: to develop a diagnostic test for OSAH from a single ECG lead.
The dataset for this challenge is divided into a learning set and a test set. Each of these sets consist of 35 recordings of the modified lead V2 of the patient’s ECG recorded during nocturnal rest.
Both training and test sets are made of 20 recordings of patients suffering from OSAH, 5 recordings of patients who were on the borderline between normality and OSAH, and 10 recordings of control patients who did not suffer from the disorder. Each recording includes minute by minute annotations indicating the presence or absence of apneas during that minute. 

The goal of the second part of the challenge, the one which will be addressed here, is to detect whether or nor the patient has suffered an apnea during each minute of nocturnal rest. 
To this end we shall use two algorithms previously published in the bibliography which are based on the calculation of a ratio between HRV spectral power in two different bands.
Specifically, we have computed the Drinnan ratio ($R_d$) \cite{drinnanDetection} and the Otero ratio ($R_o$) \cite{oteroDetection}. The Drinnan ratio and the Otero ratio are defined as $R_d=\frac{P(\left[0.005,0.01\;Hz\right])}{P(\left[0.01,0.05\;Hz\right])}$  and 
$R_o=\frac{P(\left[0.026,0.06\;Hz\right])}{P(\left[0.06,0.25\;Hz\right])},$ respectively. Both ratios were computed with RHRV using both wavelet and
Fourier analysis. 

The ratios obtained for each minute of the recording were fed to a support vector machine (SVM) \cite{RSVM}. The SVM was trained using the learning set and validated on the validation set of the Apnea ECG Database. The scores (percentage of minutes labelled correctly) obtained
in the minute by minute apnea classification using $R_d$, $R_o$ and $(R_d,R_o)$ as SVM parameters when the spectral power in the bands was calculated using wavelets were: 74.9\%,
68.4\% and 78.5\%, respectively. When using Fourier, the scores were 71.4\%, 66.8\% and 75.4\%, respectively.

Wavelet-based analysis performs slightly better than Fourier-based analysis in all scenarios. This may be due to the non-stationary nature of the signal being analyzed, and the fact that the higher temporal resolution of the wavelet analysis can help minimize the spectral contributions of apneas which have occurred outside the minute in question, but close to the end of the previous minute or to the beginning of the following one. 

\section{Discussion\label{sec:Discussion}}
We have presented an algorithm to perform HRV power spectrum analysis based on the MODWPT. The computational load of the our algorithm is comparable to the load of widely used STFT-based algorithms. The algorithm has been validated over simulated RR series with known spectral components. We have shown that the STFT and the windowed Burg method miss some quick changes that are successfully identified by the MODWPT. These results suggest that wavelet-based analysis is a better tool to analyze fast transient phenomena in the RR time series than other techniques based on windowing.

In order to obtain optimal temporal
resolution, we should avoid descending to deep levels of the PMODWPT decomposition tree. 
In RHRV, a warning is generated if the band cover needed for the analysis requires expanding more than $\log_2(N/(L - 1) + 1)$ levels,
  $N$  being the number of samples of the signal and $L$ the filter length. A
careful selection of the frequency bands to be analyzed provides
some control over the depth of the tree. For example, if the RR time
series is sampled at 4 Hz, and we want to obtain the power in the
band $[0.27, 0.5]$ Hz with a tolerance in the position of the band's
boundaries of 0.01, our algorithm will need to descend seven levels
on the tree. However, to calculate the power in $[0.26, 0.5]$ Hz with 
the same tolerance our algorithm only needs to descend three levels. 
In this way, we obtain a good estimate of the power in the band $[0.27, 0.5]$ Hz
without compromising the temporal resolution of the results.

A corollary of the phenomenon described in the previous paragraph is
that, in order to achieve optimum temporal resolution (and therefore maximize the chance of identifying fast transient phenomena), the spectral bands used in HRV analysis with wavelets will
probably have to differ from those traditionally associated with VLF, LF and HF. This opens the question of what may be the pathophysiological
significance of spectral bands different from those which have
already been widely studied in the literature.

The mother wavelet used in HRV analysis also influences the
time-frequency resolution because it determines the filter shape. Further study on how the
mother wavelet influences the results of the HRV analysis
is required. For example, our initial tests suggest that shorter wavelets have greater time
resolution than the larger wavelets, whereas the latter have a better filtering behavior
than the first ones. 

When performing a wavelet analysis, a careful choice of the bands is needed to avoid heavy computations,  as well as a choice of the mother wavelet. When using STFT analysis we also have to choose the frequency bands (although we have more flexibility in the choice), as well as the
window size, window displacement and zero padding size. Overall, more parameter choices need to be made when using the STFT. Furthermore, the selection of
the STFT parameters is complex, not only because of the computation efficiency,
 but also because the use of a certain window size influences
the temporal resolution and the frequency bands available to the analysis. To address this issue, some authors use different windows for each frequency band. This makes the analysis process heavy. Moreover, the results obtained in each of the bands are not comparable since they have been obtained using different windows. The need for tuning a lower number of knobs, and the possibility of using the same parameters in the analysis of all the HRV frequency bands are two additional advantages of using wavelets in HRV studies. 

The  algorithm described in this paper has been implemented in the RHRV package for the R environment in its 3.0 version.
To the best of our knowledge, RHRV is the first
HRV analysis toolkit that supports wavelet-based
spectral analysis of the RR time series. This software can be freely downloaded from \cite{cran}. The availability of this package will enable researchers to carry out HRV power spectrum analysis based on the wavelet transform in a simple manner (see Listing \ref{lst:basicexample}). We hope that this will help increase the number of HRV studies that use the higher temporal resolution wavelet-based techniques.

\section*{Acknowledgements}
This work was supported in part by the European Regional Development Fund (ERDF/FEDER) under the project CN2012/151 of the Galician Ministry of Education.\\

NOTICE: this is the author’s version of a work that was accepted for publication in Biomedical Signal Processing and Control. Changes resulting from the publishing process,
such as peer review, editing, corrections, structural formatting, and other quality control mechanisms may not be reflected in this document. Changes may have been made to
this work since it was submitted for publication. A definitive version was subsequently published in Biomedical Signal Processing and Control, Volume 8, Issue 6, November 2013,
Pages 542–550, DOI:10.1016/j.bspc.2013.05.006






\newpage
\textbf{Figure captions}
\begin{itemize}
\item Figure 1: MODWPT decomposition tree with the nodes selected to cover the band $[0,7/16]$ Hz. $\mathbf{W}_{0,0}$ represents the original signal, $f(t).$
\item Figure 2: Prune procedure using the PMODWPT. The crosses indicates which nodes have been pruned.
\item Figure 3: Time shift using MODWPT without (top) and with (bottom) ``Center of Energy" correction. The vertical lines indicate where spectral power should be.
\item Listing 1: Pseudocode of our wavelet-based HRV analysis algorithm.
\item Listing 2: HRV wavelet-based analysis in RHRV.
\item Figure 4: Spectrogram analysis of the simulated RR series.
\item Table 1: Relative power per frequency and per zone for STFT, windowed Burg method and wavelet analysis.
\item Figure 5: Performance of HRV analysis using PMODWPT, MODWPT and STFT.
\end{itemize}
\end{document}